\documentclass[aps,prl,twocolumn,showpacs,superscriptaddress,bibnotes]{revtex4}

\usepackage{graphicx}
\usepackage{dcolumn}
\usepackage{bm}

\begin{document}


\title{Mutual Experimental and Theoretical Validation of Bulk 
Photoemission Spectra of Sr$_{1-x}$Ca$_x$VO$_3$}


\author{A. Sekiyama}
\author{H. Fujiwara}
\author{S. Imada}
\author{S. Suga}
\affiliation{Department of Material Physics, 
Graduate School of Engineering Science, Osaka 
University, Toyonaka, Osaka 560-8531, Japan}
\author{H. Eisaki}
\altaffiliation[Present Address: ]{National Institute of Advanced Industrial 
 Science and Technology (AIST), Umezono, Tsukuba, Ibaraki 305-8568, 
 Japan}
\author{S. I. Uchida}
\affiliation{Department of Advanced Materials Science, 
Graduate School of Frontier Sciences, 
University of Tokyo, Tokyo 113-8656, Japan}
\author{K. Takegahara}
\affiliation{Department of Materials Science and Technology, 
Hirosaki University, Hirosaki, Aomori 036-8561, Japan}
\author{H. Harima}
\affiliation{The Institute of Scientific and Industrial Research, 
Osaka University, Ibaraki, Osaka 567-0047, Japan}
\author{Y. Saitoh}
\affiliation{Department of Synchrotron Research, Kansai Research 
Establishment, Japan Atomic Energy Research Institute, SPring-8, 
Mikazuki, Hyogo 679-5198, Japan}
\author{I. A. Nekrasov}
\affiliation{Institute of Metal Physics, Russian Academy of 
Sciences-Ural Division, 620219 Yekaterinburg GSP-170, Russia} 
\affiliation{Theoretical Physics III, Center for Electronic 
Correlations and Magnetism, University of Augsburg, 86135 Augsburg, 
Germany} 
\author{G. Keller}
\affiliation{Theoretical Physics III, Center for Electronic 
Correlations and Magnetism, University of Augsburg, 86135 Augsburg, 
Germany} 
\author{D. E. Kondakov}
\author{A. V. Kozhevnikov}
\affiliation{Institute of Metal Physics, Russian Academy of 
Sciences-Ural Division, 620219 Yekaterinburg GSP-170, Russia} 
\affiliation{Theoretical Physics III, Center for Electronic 
Correlations and Magnetism, University of Augsburg, 86135 Augsburg, 
Germany} 
\affiliation{Department of Theoretical Physics and Applied Materials, 
USTU, 620002 Yekaterinburg Mira 19, Russia} 
\author{Th. Pruschke} 
\affiliation{Institute for Theoretical Physics,
University of G\"ottingen, 37077 Goettingen,
Germany} 
\author{K. Held} 
\affiliation{Max Planck Institute for Solid State Research, 
Heisenbergstr.\ 1, 70569 Stuttgart, Germany}
\author{D. Vollhardt} 
\affiliation{Theoretical Physics III, Center for Electronic 
Correlations and Magnetism, University of Augsburg, 86135 Augsburg, 
Germany} 
\author{V. I. Anisimov} 
\affiliation{Institute of Metal Physics, Russian Academy of 
Sciences-Ural Division, 620219 Yekaterinburg GSP-170, Russia} 
\affiliation{Theoretical Physics III, Center for Electronic 
Correlations and Magnetism, University of Augsburg, 86135 Augsburg, 
Germany}


\date{\today}

\begin{abstract}
We report high-resolution high-energy photoemission spectra 
together with parameter-free LDA+DMFT (local density approximation 
+ dynamical mean-field theory) results for Sr$_{1-x}$Ca$_x$VO$_3$, 
a prototype 3$d^1$ system. In contrast to earlier investigations 
the bulk spectra are found to be insensitive to $x$. 
The good agreement between experiment and theory confirms 
the bulk-sensitivity of the high-energy photoemission spectra.
\end{abstract}

\pacs{79.60.Bm, 71.27.+a, 71.20.Be}

\maketitle

Electronic correlations in
transition metal oxides (TMO) are at the center of
present solid state research.  Among TMO, 
cubic perovskites have the simplest crystal 
structure and, thus, may be
 viewed as a starting point for understanding 
the electronic properties of more complex systems. 
Typically, the 3$d$ states in those materials form comparatively 
narrow bands of width $W\!\approx\!2\!-\!3$ eV which lead to strong Coulomb 
correlations between the electrons. 
Particularly simple are TMO with a 3$d^1$ configuration 
like Sr$_{1-x}$Ca$_x$VO$_3$
since they do not show a complicated electronic structure. 

Intensive experimental investigations of spectral and transport 
properties of strongly correlated 3$d^1$ TMO started 
with the paper by Fujimori {\it et al.}~\cite{AFSCVO} 
These authors observed a pronounced lower Hubbard band  in the photoemission spectra (PES) measured at low photon energies
($h\nu\! \leq\! 120\,$ eV) which cannot be explained by conventional band 
structure theory. 
A number of papers~\cite{IHI95,Morikawa} subsequently addressed 
the spectral, transport and thermodynamic properties of 
Sr$_{1-x}$Ca$_x$VO$_3$, yielding contradictory results. 
While the thermodynamic and transport properties (Sommerfeld 
coefficient, resistivity, and paramagnetic susceptibility) 
do not change much with $x$, low photon energy (low-$h \nu$) PES show 
drastic differences for varying $x$. 
In fact, these spectroscopic data seemed to imply that 
Sr$_{1-x}$Ca$_x$VO$_3$ is on the verge of a Mott-Hubbard transition 
for $x\!\to\! 1$.
This is in concordance with the widespread theoretical expectation
that the bandwidth  $W$ should decrease strongly with increasing $x$
since the V-O-V angle is reduced from
$\theta = 180^{\circ}$ for cubic SrVO$_3$~\cite{Rey} to 
$\theta = 154.3^{\circ}$ (rotation angle), 171$^{\circ}$ 
(titling angle)~\cite{MHJung} for orthorhombic 
CaVO$_3$.~\cite{Chamberland}

With the SPring-8 beamline we are now able to reexamine this
puzzling discrepancy by high-resolution
high-energy
PES which, owing to a longer  photoelectron mean free path 
$\lambda$ at $h \nu \approx 1000\,$eV, is much more 
bulk-sensitive.~\cite{Tanuma,ASN,RSI,MaitiPRL,MaitiEU} 
In this Letter, we show that 
these bulk-sensitive PES 
on fractured surfaces  with $\approx 100\,$meV resolution 
are nearly independent of the 
Ca concentration $x$ and significantly different from 
high-energy PES at 0.5$\,$eV resolution on scraped surfaces.~\cite{MaitiEU}
The good agreement with
 LDA+DMFT~\cite{Metzner,VIA97,AL2000,IAN2000,AIL2001,Savrasov,Psik} 
spectra provides an important cross-validation: By means of a
TMO prototype we show that the two
cutting-edge methods employed here are indeed able to determine bulk spectra.

\begin{figure}
\includegraphics[width=8.5cm,clip]{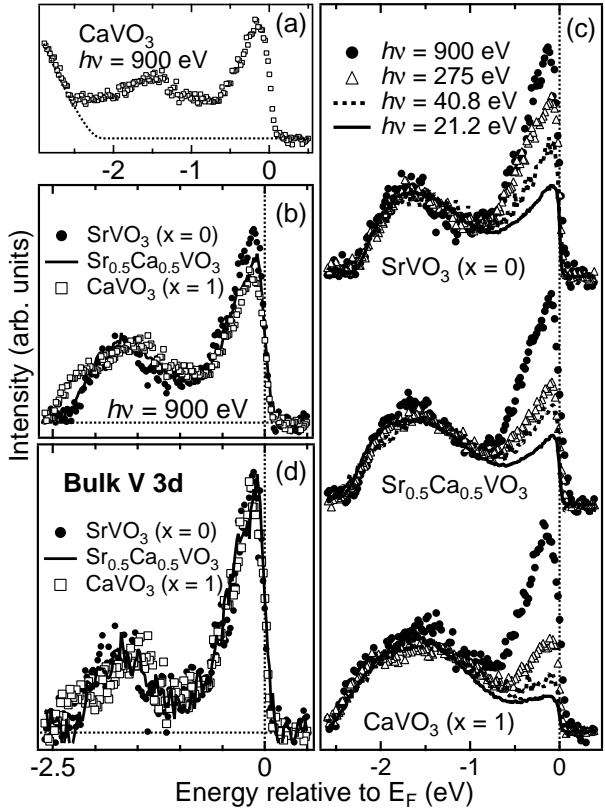}%
\caption{\label{Fig1}(a) Raw spectrum of the high-resolution PES 
near $E_F$ of CaVO$_3$ (squares) and the fitted tails of O $2p$ 
contributions. (b) V $3d$ PES of Sr$_{1-x}$Ca$_x$VO$_3$ 
at $h\nu =$ 900 eV obtained by subtracting the fitted tails of O $2p$ 
contributions from the raw spectra as in (a). 
(c) $h\nu$ dependence of the V $3d$ PES normalized by 
the incoherent spectral weight ranging from 
$-0.8$ to $-2.6$ eV. (d) Bulk V $3d$ PES of Sr$_{1-x}$Ca$_x$VO$_3$ 
as obtained by the procedure described in the text.}
\end{figure}

The high-resolution PES  are summarized 
in Fig.~\ref{Fig1}.~\cite{ExpCond} 
In all spectra, the peak near $E_F$ and 
the broad peak centered at about $-1.6$ eV correspond to 
the coherent (quasiparticle peak) and 
incoherent parts (lower Hubbard band), respectively. 
In contrast to low-$h \nu$ PES 
of Sr$_{1-x}$Ca$_x$VO$_3$ reported so far, including our own,
the spectral variation with $x$ is noticeably smaller 
in the spectra measured at $h\nu =$ 900 eV as summarized 
in Fig.~\ref{Fig1}(b). 
As shown in Fig.~\ref{Fig1}(c), the coherent spectral weight 
decreases drastically with decreasing $h\nu$ for all compounds. 
One might consider that the relatively strong incoherent spectral 
weight of the $-1.6$ eV peak at low $h\nu$ originates possibly from a 
large O $2p$ weight in this peak. 
However, both coherent and incoherent parts are drastically enhanced 
in a V $2p-3d$ resonance photoemission (not shown), 
and are therefore assigned to the effective V $3d$ states
 around $E_F$.~\cite{O2p} 
Since the oxygen admixture to these predominantly $3d$ states
is very minor,  the  relative V $3d$/O $2p$ cross section which
also changes by only 8\% from $h\nu =$ 275 and 900 eV~\cite{Lindau}
can be neglected. Also note that the effect of
PES matrix elements is  generally small for high-$h\nu$
angle-integrated spectra. 
Therefore, the monotonous increase of the coherent part with $h\nu$ 
is inevitably attributed to an increased bulk sensitivity.

We have obtained the bulk V $3d$\cite{O2p}  PES, Figure \ref{Fig1}(d), 
of Sr$_{1-x}$Ca$_x$VO$_3$ from the data at $h\nu =$ 
900 and 275 eV by the following procedure~\cite{HeII}: 
(1) The mean free path $\lambda$ has been calculated as $\sim$17 and 
$\sim$7 {\AA} at $h\nu =$ 900 and 275 eV.~\cite{Tanuma} 
(2) The bulk weight $R$ ($< 1$, depending on $h\nu$) should be 
determined as $\exp (-s/\lambda )$ where $s$ is a "surface thickness". 
Therefore $R$s at 900 eV ($R_{900}$) and 275 eV ($R_{275}$) 
are related as $R_{275} = {R_{900}}^{2.4}$. 
(3) The observed V $3d$ PES intensity
 at $h\nu =$ 900 eV is represented as 
$I_{900}(E) = I_B(E)R_{900} + I_S(E)(1-R_{900})$ 
while 
$I_{275}(E) = I_B(E){R_{900}}^{2.4} + I_S(E)(1-{R_{900}}^{2.4})$, 
where $I_B(E)$/$I_S(E)$ is the bulk/surface $3d$ PES 
and $E$ stands for the binding energy. 
(4) If $s$ is assumed to be 7.5 {\AA} corresponding to about twice the 
V-O-V distance,~\cite{IHI95} $R_{900}$ ($R_{275}$) is determined as 
$\sim$0.64 ($\sim$0.34). 
With this $R_{900}$ we obtain the  $3d$ bulk PES $I_B(E)$
shown in Fig.~\ref{Fig1}(d), noting that   $I_B(E)$  hardly 
changes ($< 15\%$ at $E_F$) even when we assume $s$ to 
range from 5.4 to 11 \AA.
In contrast to the previous PES, the bulk $3d$ PES 
are almost equivalent among the three compounds, 
indicating that the V-O-V distortion does not much influence 
the occupied bulk $3d$ states in Sr$_{1-x}$Ca$_x$VO$_3$. 
This is consistent with the thermodynamic properties. 
It should be noticed that on scraped sample surfaces 
nearly $x$-independent bulk spectral functions could 
not be obtained.~\cite{MaitiEU}

\begin{figure}
\includegraphics[width=7.5cm,clip]{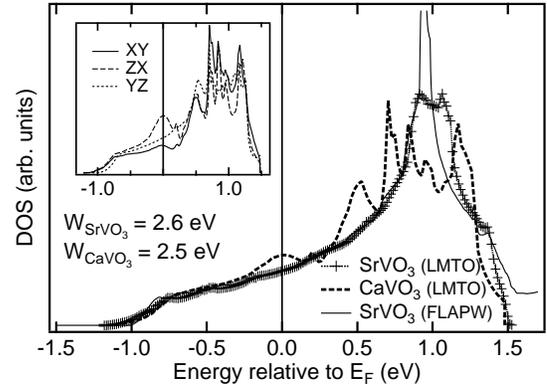}%
\caption{\label{Fig2}Comparison of the LDA DOS of the V-$3d$ $t_{2g}$ 
band for SrVO$_3$ (line with cross marks) and CaVO$_3$ (dashed line) 
obtained by 
using the LMTO method. The V $3d$ ($t_{2g}+e_g$) partial DOS for 
SrVO$_3$ (solid line) obtained by using the FLAPW method is also shown. 
In CaVO$_3$ the degeneracy of the three $t_{2g}$ bands is lifted 
(see inset).}
\end{figure}

Figure~\ref{Fig2} shows the LDA density of states (DOS) for 
SrVO$_3$ and CaVO$_3$ which we obtained using the TBLMTO47 code of 
Andersen and coworkers~\cite{LMTO} as well as a full-potential 
linearized augmented plane wave (FLAPW). 
The van-Hove-like peak of the latter can also be obtained by 
LMTO,~\cite{Pavarini} leading to minor differences which are, 
however, not important for our LDA+DMFT calculations. 
Most importantly, the one-electron $t_{2g}$ bandwidth of CaVO$_3$, 
defined as the energy interval where the DOS in Fig.~\ref{Fig2} 
is non-zero, is found to be only 4\% smaller than that of SrVO$_3$ 
($W_{\mbox{CaVO}_3}$ = 2.5 eV, $W_{\mbox{SrVO}_3}$ = 2.6 eV). 
The small reduction of the bandwidth 
well agrees with \cite{Pavarini} and
will later turn out to be important.

\begin{figure}
\includegraphics[width=8.5cm,clip]{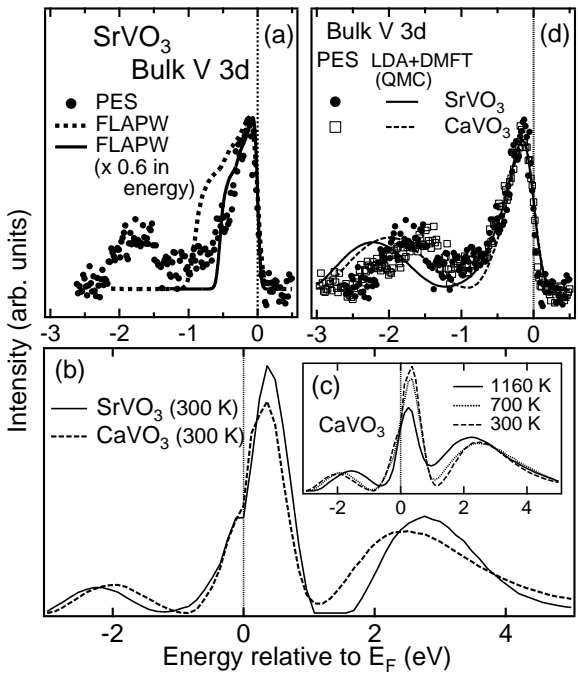}%
\caption{\label{Fig3}(a) Comparison of the bulk $3d$~\cite{O2p} PES of SrVO$_3$ 
(dots) to the $3d$ DOS obtained by using the 
FLAPW method (dashed curve), which has been multiplied with the 
Fermi function at 20 K and then broadened by the 
experimental resolution of 140 meV. The solid curve shows the same 
$3d$ DOS but the energy is scaled down by a factor of 0.6. 
(b) LDA+DMFT (QMC) spectrum of SrVO$_3$ (solid line) and CaVO$_3$ 
(dashed line) calculated at 300 K.
(c) Effect of the temperature in the case of CaVO$_3$. 
(d) Comparison of the bulk PES with the calculated LDA+DMFT spectra 
for SrVO$_3$ and CaVO$_3$. }
\end{figure}

In order to determine the genuine effect of the electron correlations
on the bulk spectra, 
we first compare in Fig.~\ref{Fig3}(a) the measured spectrum with a V $3d$
partial DOS for SrVO$_3$ obtained by using the FLAPW method.
Here, the partial DOS broadened by the instrumental resolution 
(dashed curve) is normalized to the bulk $3d$ spectral function of 
SrVO$_3$ by the integrated intensity from $E_F$ to $-2.6$ eV. 
This comparison shows  that the band-structure calculation does not
reproduce the incoherent spectral weight at all and that 
the width of the observed coherent part is about 
60\% of the predicted value. 
Clearly, these deviations originate from electronic correlations 
which are not treated correctly in the band-structure 
calculation. 

For a more quantitative analysis, we use LDA+DMFT as a non-perturbative 
approach,~\cite{VIA97,Psik,Pavarini} in which the local intra-orbital Coulomb 
repulsion $U$, the inter-orbital repulsion $U'$, and the exchange 
interaction $J$ are explicitly taken into account. 
We calculated these interaction strengths by means of the constrained 
LDA method~\cite{OG89} for SrVO$_3$. The resulting value of the 
averaged Coulomb interaction is $\bar{U}$ = 3.55 eV ($\bar{U} = U'$ 
for $t_{2g}$ orbitals~\cite{Psik,Zolfl}) and $J$ = 1.0 eV. 
$U$ is then fixed by the rotational invariance to $U = U' + 2J =$ 
5.55 eV. 
These calculated values are in agreement with previous calculations 
of vanadium systems~\cite{Solovyev} and experiments.~\cite{Makino} 
We use these values in our LDA+DMFT calculations 
for both SrVO$_3$ and CaVO$_3$~\cite{UJCa} and 
employ quantum 
Monte Carlo (QMC)~\cite{QMC} simulations to solve 
the DMFT self-consistency equation.
The resulting LDA+DMFT (QMC) spectra for SrVO$_3$ and CaVO$_3$ 
in Fig.~\ref{Fig3}(b) show genuine correlation effects, i.e., 
the formation of lower Hubbard bands at about $-2$ eV and upper 
Hubbard bands at about $2.5$ eV, with well pronounced 
quasiparticle peaks (coherent part) near $E_F$. 
The results indicate that both SrVO$_3$ and CaVO$_3$ are strongly 
correlated metals, but not on the verge of a Mott-Hubbard 
metal-insulator transition. 
The 4\% difference in the LDA bandwidth between the compounds 
is only reflected in some additional transfer of spectral 
weight from the coherent to incoherent parts. 
The many-particle spectra of the two systems are seen to be quite 
similar, especially in the occupied states, 
in agreement with Ref.~\cite{Pavarini}. 
Figure~\ref{Fig3}(c) shows that the effect of temperature 
on the spectrum is small for $T\lesssim$ 700 K. 
The slight differences in the quasiparticle peaks between SrVO$_3$ 
and CaVO$_3$ lead to different effective masses, 
namely $m^*/m_0 = 2.1$ for SrVO$_3$ and 2.4 for CaVO$_3$, 
which agree with $m^*/m_0 = 2-3$ for SrVO$_3$ 
and CaVO$_3$ as obtained from thermodynamics~\cite{IHI95} and 
de Haas-van Alphen experiments.~\cite{IHI02} 

In Figure~\ref{Fig3}(d), we compare 
the experimental bulk V $3d$~\cite{O2p} PES to the LDA+DMFT spectra (300 K), 
which were multiplied with the Fermi function at the 
experimental temperature (20 K) and broadened with the PES resolution. 
The quasiparticle peaks in theory and experiment are seen to be in 
very good agreement. 
In particular, their height and width are almost identical for 
both SrVO$_3$ and CaVO$_3$.
There is some difference in the positions of the incoherent part which
may be partly 
due to uncertainties in the {\it ab-initio} calculations of $\bar{U}$
or in the oxygen subtraction procedure of Fig.\ \ref{Fig1}.
Indeed, some differences have to be expected since both
the experimental determination of the bulk spectrum and the LDA+DMFT 
method involve approximations. 
Hence, our good overall agreement confirms that
the new high-energy high-resolution 
PES indeed measures bulk spectra which 
can be calculated  quite accurately by 
LDA+DMFT without free-parameters.

This work was supported by a Grant-in-Aid for COE Research 
from the Ministry of Education, Culture, Sports, Science 
and Technology (MEXT), Japan, by the Deutsche Forschungsgemeinschaft 
through SFB 484 and the Emmy-Noether program, the Russian Foundation 
for Basic Research through RFFI-01-02-17063 and RFFI-03-02-06126, 
the Ural Branch of the Russian Academy of Sciences for Young 
Scientists, Grant of the President of Russia MK-95.2003.02, and the 
National Science Foundation under Grant No. PHY99-07949. 
The photoemission at $h\nu =$ 275 and 900 eV was performed 
under the approval of the Japan Synchrotron Radiation Research 
Institute (1999B0076-NS-np).
We thank T. Ushida, A. Shigemoto, T. Satonaka, T. Iwasaki, M. Okazaki, 
S. Kasai, A. Higashiya, K. Konoike, A. Yamasaki, A. Irizawa 
and the staff of SPring-8, especially T. Muro, T. Matsushita 
and T. Nakatani for supporting the experiments. 
We also thank A. Sandvik for making available his maximum entropy 
code and acknowledge valuable discussions with I. H. Inoue, 
A. Fujimori, R. Claessen, and M. Mayr. 
\end{document}